\documentclass[ reprint, superscriptaddress, amsmath,amssymb, aps, prb, floatfix]{revtex4-2}

\usepackage{graphicx}
\usepackage{dcolumn}
\usepackage{bm}
\usepackage{xcolor}
\usepackage{hyperref}
\usepackage[mathlines]{lineno}

\begin{document}

\title{Unidirectional spin waves measured using propagating spin wave spectroscopy}
\author{G. Y. Thiancourt}
\thanks{}

\author{S. M. Ngom}
\author{N. Bardou}
\affiliation{%
Université Paris-Saclay, CNRS, Centre de Nanosciences et de Nanotechnologies, 91120, Palaiseau, France
}%

\author{T. Devolder}%
\email{thibaut.devolder@universite-paris-saclay.fr}
\affiliation{%
Université Paris-Saclay, CNRS, Centre de Nanosciences et de Nanotechnologies, 91120, Palaiseau, France
}%

\date{\today}

\begin{abstract}
The dispersion relation of spin waves can vary monotonically about the center of the Brillouin zone, allowing zero-momentum wavepackets to flow unidirectionally, which is of interest for applications. Techniques such as propagating spin wave spectroscopy are inoperative in such cases because of the difficulty to identify the spin wave wavevector at a particular frequency within a spectrum. Here we present a method to analyse this case and apply it to acoustic spin waves in a synthetic antiferromagnet in the scissors state, in which we confirm that propagation parallel to the applied fields is unidirectional. Interestingly, we find that in this unidirectional situation, the phase accumulated by the spin waves propagating between two antenna is not proportional to the antenna spacing. It is also a function of the two other lengths of the problem: the antenna width and the spin wave decay length. Accounting for them is required to avoid wavevector errors in the dispersion relations.
\end{abstract}
\maketitle

Thin film magnetism finds versatile applications in the radio-frequency domain, where understanding spin wave (SW) dynamics is crucial. The emergence of Propagating Spin Wave Spectroscopy (PSWS \cite{bailleul_propagating_2003, vlaminck_current-induced_2008, ciubotaru_all_2016, talmelli_reconfigurable_2020}) has garnered attention due to its swift, precise, and sensitive capabilities, making it a useful method for investigating SW propagation on the micron scale \cite{yu_magnetic_2014, che_efficient_2020}, including the assessment of their group velocities \cite{ciubotaru_all_2016, collet_spin-wave_2017, vanatka_spin-wave_2021}. While utilizing PSWS for determining SW band structure holds promise, it necessitates the precise disentanglement of each band's contribution to the overall PSWS signal\cite{devolder_measuring_2021}. Often, this entails employing semi-empirical models for data analysis  \cite{maendl_spin_2017, wang_chiral_2020, sushruth_electrical_2020, vanatka_spin-wave_2021}, which may yield ambiguous outcomes. Achieving a robust resolution of dispersion relations for different SW modes typically mandates either multiple variants of a sample \cite{vanatka_spin-wave_2021} or the external measurement of ferromagnetic resonance \cite{devolder_measuring_2021}. Notably, the scenarios investigated using PSWS thus far have predominantly been reciprocal or weakly non-reciprocal; the development of precise PSWS methodologies for strongly non-reciprocal situations \cite{kwon_giant_2016, chen_excitation_2019, temdie_high_2023} remains a frontier yet to be explored. 

In this study, we implement PSWS on a synthetic antiferromagnet (SAF) whose dispersion relation $\omega(k)$ has been shown to be highly non-reciprocal \cite{ishibashi_switchable_2020, gallardo_reconfigurable_2019} and even monotonic \cite{millo_unidirectionality_2023} across $k=0$ when the wavevector is parallel to the applied field.
Under these conditions, the group velocity remains positive regardless of the wavevector's sign, resulting in a unidirectional energy flow of spin waves.
This paper builds upon the PSWS analysis from ref.~\onlinecite{devolder_propagating-spin-wave_2023} by applying it to SAF spin waves. The methodology incorporates field-differentiation, time-of-flight filtering and precise analysis of PSWS signal phases. 
Notably, our study reveals that the existence of several characteristic lengths --antenna width $w_\textrm{ant}$, antenna-to-antenna distance $r$ and SW attenuation length $L_\textrm{att}$-- entails that the phase accumulated by the SW during their journey between antennas is not directly proportional to $r$: The commonly done assumption that the propagation distance is the antenna-to-antenna distance does not apply when the dispersion relation is monotonic. The correct analysis of the PSWS signal allows to demonstrate that the unidirectional energy flow for acoustic SWs in SAF occurs across a broad range of applied fields.

\section{\label{sec:Propagating Spin Wave Spectroscopy procedure}Propagating Spin Wave Spectroscopy measurement procedure}
\begin{figure}
\centering
	\includegraphics[width=0.45\textwidth]{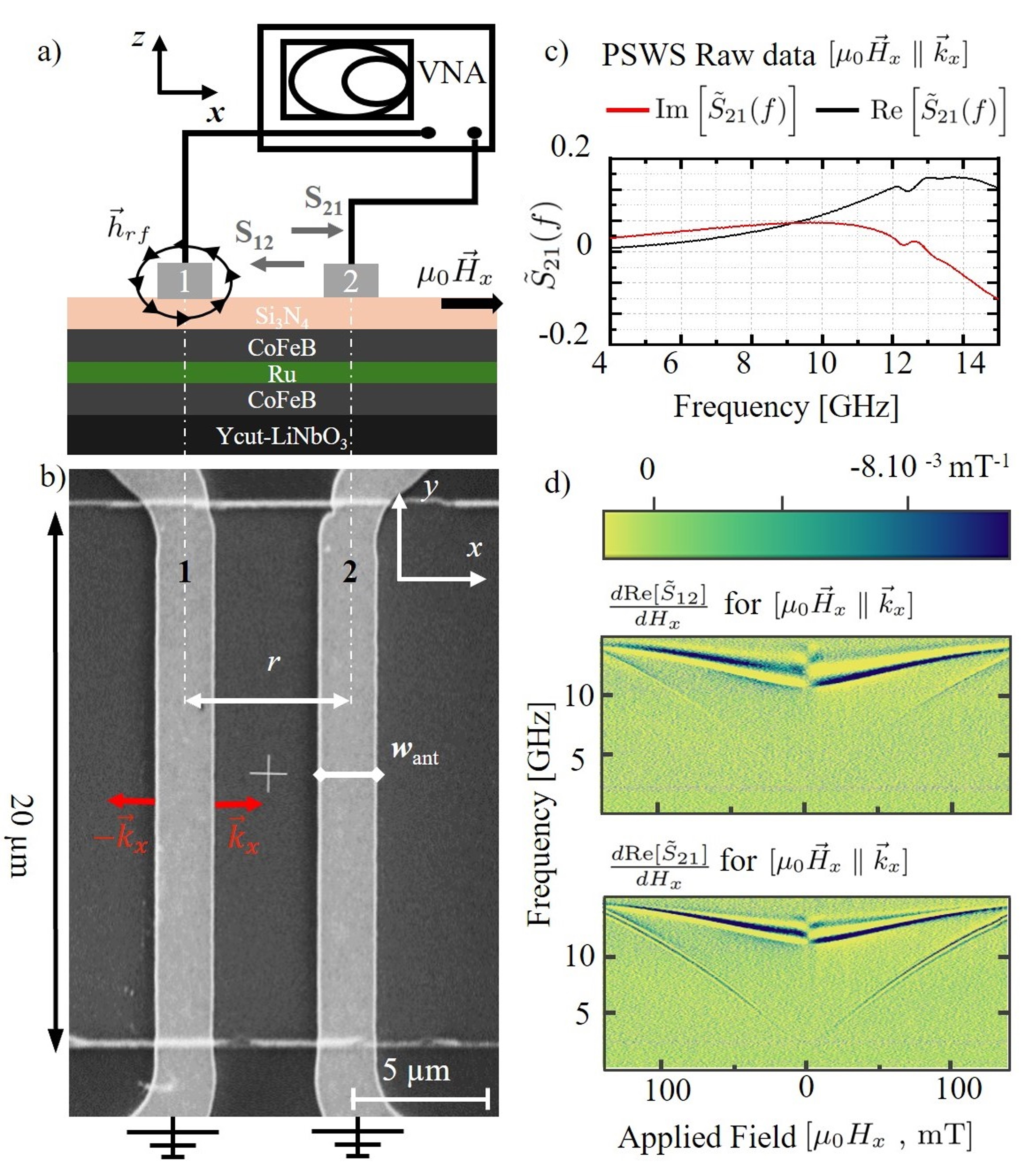}
	\caption{\label{fig:geometry} 
		 Propagating spin waves spectroscopy (PSWS) setup and measurements for device with $\{w_\textrm{ant}, r\}=\{2~\mu\textrm{m}, ~4~\mu\textrm{m}\}$. (a) Schematic of the experimental setup with $\mu_0 \vec{H}_x$ the applied field and $\vec{h}_{rf}$ the exciting field of antenna $1$. 
		(b) Scanning electron micrograph of a device. Light gray: single-wire antennas with a width $w_\textrm{ant}$. Medium gray: Synthetic antiferromagnet conduit for SWs. (c) Forward transmission parameter $\tilde{S}_{21}$ for $\mu_0 H_x = 93$ mT. (d) Field derivative of $\text{Re}[\tilde S_{ij}]$ versus frequency and field.
	} 
	%
\end{figure}

 We studied SAF films with the composition LiNbO$_3$ (substrate)/Ta/CoFeB($t_\textrm{mag}~=~17$ nm)/ Ru(0.7 nm) /CoFeB($t_\textrm{mag}$)/Ru /Ta (cap) (Fig \ref{fig:geometry}.a, growth described in ref.~\onlinecite{mouhoub_exchange_2023}). Properties include a CoFeB magnetization of $\mu_0M_s = 1.7~ \textrm T$, and an interlayer exchange energy $J_{\text{ex}} = -1.7~\textrm{mJ}/\textrm{m}^2$, corresponding to an interlayer exchange field $\mu_0H_j = - {2J}/{ (M_s t_{\textrm{mag})}}=148~\textrm{mT}$. Under applied fields in the 0-100 mT range, the linewidth of the acoustic spin wave resonance at $k=0$ is almost constant at $\Delta\omega_0 /{(2 \pi)}~=~ 350 \pm 50$ MHz, which is consistent with a Gilbert damping $\alpha = \Delta \omega_0 / (\gamma \mu_0 (M_s + H_j))= 0.006 \pm 0.001$, where $\gamma$ is the gyromagnetic ratio. 

To perform electrical PSWS, the films are patterned into devices (Fig \ref{fig:geometry}.a) consisting of $20~\mu \textrm{m}$-wide SW conduits inductively coupled to two identical single-wire antennas. The antenna widths are $w_\textrm{ant}=1,~2$ and $2~\mu\textrm m$ and they are respectively spaced at center-to-center distances $r=3.4, ~4$ and $6~\mu\textrm m$ (Fig \ref{fig:geometry}.b). 

The two antennas are connected to a two-port Vector Network Analyzer (VNA) in order to measure the forward ($\tilde{S}_{21}$) and backward ($\tilde{S}_{12}$) transmission parameters in the presence of a static field ${H}_x$ applied along the SW propagation path (see Fig \ref{fig:geometry}.a). 
The insertion loss, i.e., $20\,\textrm{log}_{10}(||\tilde{S}_{21}||)$, is typically -24.7 dB and -16 dB at 5 and 10 GHz respectively, both for $\{w_\textrm{ant}, r\}=\{2~\mu\textrm{m}, ~4~\mu\textrm{m}\}$.
A representative transmission spectrum is shown in Fig.~\ref{fig:geometry}.c. 
A substantial part of the transmission signal arises unfortunately from the antenna-to-antenna cross-talk that is not related to SWs. 

Two methods can reveal the SW contributions to the transmission parameters. A reference signal $\tilde S_{ij}^\textrm{ref}$ can first be subtracted from the raw data. Here we use the $H_x$-averaged transmission coefficients $\tilde S_{ij}^\textrm{ref}=\langle \tilde S_{ij} \rangle_{H_x}$. The field-dependent part of the  transmission is defined as: 
\begin{equation} 
\label{eq:Sbar}
\overline S_{ij}(H_{x}, \omega)=\tilde S_{ij}(H_{x}, \omega) - \tilde S_{ij}^\textrm{ref}(\omega) .
\end{equation}
Since the reference signal unavoidably includes some SW contribution, Eq.~\ref{eq:Sbar} is an imperfect suppression of the SW-independent background. 

\begin{figure}
\centering
    \includegraphics[width=0.5\textwidth]{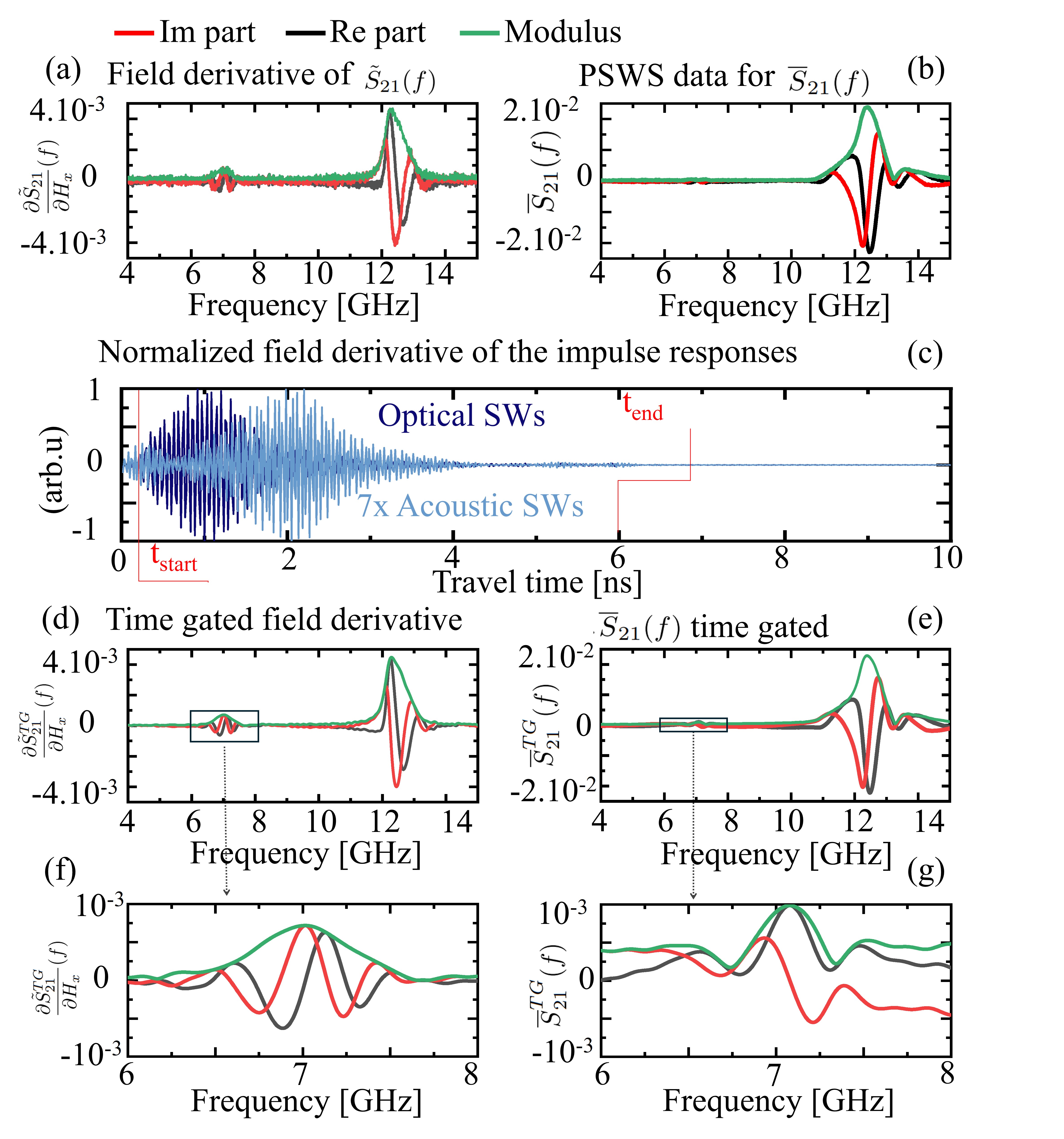}
\caption{\label{fig:signal_processing} 
Spin wave spectra and their transformation using our data processing methods for $\mu_0 H_x=-50~\textrm{mT}$, $r=4~\mu \textrm{m}$ and $w=2 ~\mu$m. (a) Field derivative of $\tilde S_{21}$.(b) Field dependent part of $\tilde S_{21}$. (c) Spinwave wavepacket in time domain. The rectangular window of the time gating is defined at $t_{\textrm{start}}=0.025~$ns and ending at $t_{\textrm{end}}=6~$ns. (d-e) Time gated versions of panels (a) and (b). (f-g) Zooms of the contribution of the acoustic spin wave mode.}

\end{figure}

Alternatively, if the magnetic susceptibility varies with the applied field, the contribution of the SWs can also be determined from the field derivative $\frac{d \tilde S_{ij}}{d H_x}$ of the transmission signals. This differentiation effectively enhances the SW-related oscillation in the transmission signals, however at the expense of increased noise [compare Fig. \ref{fig:signal_processing}.(a) and (b)].

The goal here is to accurately determine the SW dispersion relations $\omega(k)$ of our SAF from PSWS data. This requires a detailed analysis of the phase of the transmission coefficients \cite{ciubotaru_all_2016, devolder_propagating-spin-wave_2023}; noise and background should thus be minimized. Time-of-flight spectroscopy using time-gating of the PSWS data \cite{devolder_measuring_2021} can be used as an supplemental processing step for that purpose. This method consists of the mathematical calculation of the impulse response $s_{21}(t)$ from the measured $\tilde S_{21}(f)$ parameter, followed by the selection of the wavepackets of interest by time-of-flight filtering and truncation. Then, a back Fourier transformation is applied on the truncated impulse responses to recover a spectrum free of the noise that arose from non-spin wave contributions \cite{devolder_measuring_2021}.

Fig.~\ref{fig:signal_processing} details the time-gating procedure for a representative transmission spectrum and its field derivative. 
For qualitative understanding of the group velocities of the optical and acoustic SW branches, it is interesting to display separately the impulse responses corresponding to the sole high frequency part of the spectrum, and to the sole low frequency part of the spectrum. The wavepacket of the optical SWs (blue curve in Fig.~\ref{fig:signal_processing}c) arrives at a group delay $t_\textrm{g}$ of approximately 1 ns. The wavepacket of the slower acoustic SWs (cyan curve in Fig.~\ref{fig:signal_processing}c) arrives later at a longer group delay of circa 2 ns. A rough estimate of the group velocities is $v_g \approx r/ t_\textrm{g}$, which amount to $\approx 4~\textrm{km.s}^{-1}$ for the optical SWs and to $\approx 2~\textrm{km.s}^{-1}$ for the acoustic SW. 

 A time gate from $t_{\textrm{start}}=0.25$ ns to $t_{\textrm{end}}=6$ ns is suitable to retain the two SW wavepackets while excluding both the long-delay part of the white noise, and the short-delay antenna-to-antenna electromagnetic cross-talk [see Fig.~\ref{fig:signal_processing}(b)]. 
 The back transformation to frequency domain yields field-derivative spectra that are free of background and less noisy [compare Fig.~\ref{fig:signal_processing}(a) and (f)]. The phase of such signals can be accurately defined even on the low signal of the acoustic SW branch. The background suppression is more efficient when using the field derivative
 instead of Eq.~\ref{eq:Sbar} [compare Fig.~\ref{fig:signal_processing}(f) and (g)] which gives artifacts such as phase distortions, as well as an oscillatory modulus that is unphysical for single-wire antennas \cite{sushruth_electrical_2020}. In what follows we combine field-differentiation and time-gating when analysing the smallest SW signals.

\section{Model of propagating spin wave spectroscopy}
The general formalism relating PSWS data to SW dispersion relations is reviewed in ref.~\cite{devolder_propagating-spin-wave_2023}. We use the same notations. For an ultrathin antenna in close vicinity with the magnetic film, if the dispersion relation is \textit{monotonic} with positive slope [labeled as "/" to mimic a linear increase of $\omega(k)$], the experimental transmission parameter $\tilde S_{21}$ for \(|r|>w_{\textrm{ant}}\) has a spectral content identical to the term (Eq.~22 of ref.~\cite{devolder_propagating-spin-wave_2023}):
\begin{equation}
\label{S21} \tilde{S}_{21f}^{yy}(\omega, r> w_\textrm{ant}, /) ~~\propto ~\mathcal{L}(\omega)   P_1(r, \omega) U(r,\omega) ~,\end{equation}
where $\mathcal{L}(\omega) \propto       { 
   \left({{} (\omega -{\omega_0})}-i{v_g}/ {L_\textrm{att}} \right)^{-2}}$ is a Lorentzian centered at $\omega_0= \omega(k=0)$ 
 with a full width at half maximum $\Delta \omega_0 = { 2 v_g}/{L_\textrm{att}} $, which is the ratio of the group velocity to the attenuation length\footnote{We shall neglect any variation of the attenuation length with the wavevector.}.
 $P_1(r, \omega) = e^{i\frac{ r (\omega -{\omega_0})}{v_g}} e^{-\frac{r}{L_\textrm{att}}} $ is the propagating factor with a phase rotation $\propto r$ and an exponential decay in space.
$U(r,\omega)$ is a unidirectional term that also accounts for the finite antenna width. For  \(r>w_{\textrm{ant}}\), it reads\cite{devolder_propagating-spin-wave_2023}: 
$$U(r, \omega) = -4 
 +  2 \left( e^{\frac{w}{L_\textrm{att}} + \frac{i w (\omega -{\omega_0})}{v_g}} + e^{\frac{-i w (\omega -{\omega_0})}{v_g} -\frac{w}{L_\textrm{att}}}  \right) \label{Uder}.  $$

The experimental datasets that are the most adequate for reliable analysis are the time-gated versions of the $\frac{\partial \tilde{S}_{ij}}{\partial H_x} (\omega)$ spectra. We can deduce the dispersion relation from these spectra by relying on the following three observations. 

Firstly, $ 
    \frac{\partial \tilde{S}_{21}}{\partial H_x} (\omega) = \frac{d \omega}{d H_x}
 \frac{\partial \tilde{S}_{21}}{\partial  \omega} $ and since $\frac{d \omega}{d H_x}$ is a real number, the phase of field-derivatives and frequency-derivatives are equivalent for $\frac{d \omega}{d H_x} >0$. Moreover, $\frac{d \omega}{d H_x}$ is almost constant for the acoustic SW of a SAF ~\cite{devolder_measuring_2022}; so the modulus of field-derivatives and of frequency-derivatives are equivalent for practical purposes. \\

 Second, the modulus of derivative-spectra obey \footnote{This holds up to fourth order in $\Delta\omega_0 / \omega_0$, hence in all practical situations of PSWS.}: 
 \begin{equation}
     \Big\lvert\Big\lvert   \frac{\partial \tilde{S}_{21f}^{yy}}{\partial H_x} (\omega) \Big\rvert\Big\rvert \textrm{~has~a~maximum~at~} \omega = \omega_0 \label{modulus}
 \end{equation} 
This property is useful to determine the frequency $\omega_0/ (2 \pi)$ of the acoustic SW at $k=0$. 
 
Third, one can show from Eq.~\ref{S21} that 
\begin{equation}
    \begin{split}   & \textrm{Arg} \left( \frac{\partial \tilde{S}_{21f}^{yy}}{\partial H_x} (\omega) \right) +  \frac \pi 2 + 2\pi \mathbb Z = \\ & k \left( r + \frac{w_\textrm{ant}^2}{6 L_\textrm{att} } + \frac{w_\textrm{ant}^2}{6 r }\right)  + O(w_\textrm{ant}^4). \label{phase} \end{split}
\end{equation}
The \(2 \pi\mathbb Z\) uncertainty within Eq.~\ref{phase} can be resolved by substituting in any known point along the dispersion relation. This known point can be simply  $k=0$ and the frequency $\omega_0$ determined from Eq.~\ref{modulus}.

It is worth emphasizing the Eq.~\ref{phase} entails that when harnessing a monotonic dispersion relation, the phase of a spectrum rotates at a pace that is \textit{faster} than simply $k r$. This is in stark contrast with the usual assumption that the variation of is exactly equal to $k r$ \cite{maendl_spin_2017, wang_chiral_2020, vanatka_spin-wave_2021, devolder_measuring_2021}).  Eq.~\ref{phase} means that the \textit{effective} propagation distance is not $r$ but 
\begin{equation}
r_\textrm{eff}=\left( r + \frac{w_\textrm{ant}^2}{6 L_\textrm{att} } + \frac{w_\textrm{ant}^2}{6 r }\right) \in [r, r+ w_\textrm{ant}]. \label{reff}    
\end{equation}
 In situations of interest for PSWS, we have narrow antennas satisfying $w_\textrm{ant} \ll 6 L_\textrm{att}$ such that $r_\textrm{eff} \in [r, r+ w_\textrm{ant}]$. This recalls that while the center-to-center distance $r$ is the intuitive (hence often chosen) propagation distance, SWs can in fact also be emitted and collected at the outer edges of the antennas, which are separated by the distance $r+ w_\textrm{ant}$.

\section{Dispersion relations of acoustic spin waves}

\begin{figure}
    \centering
    \includegraphics[width=0.45\textwidth]{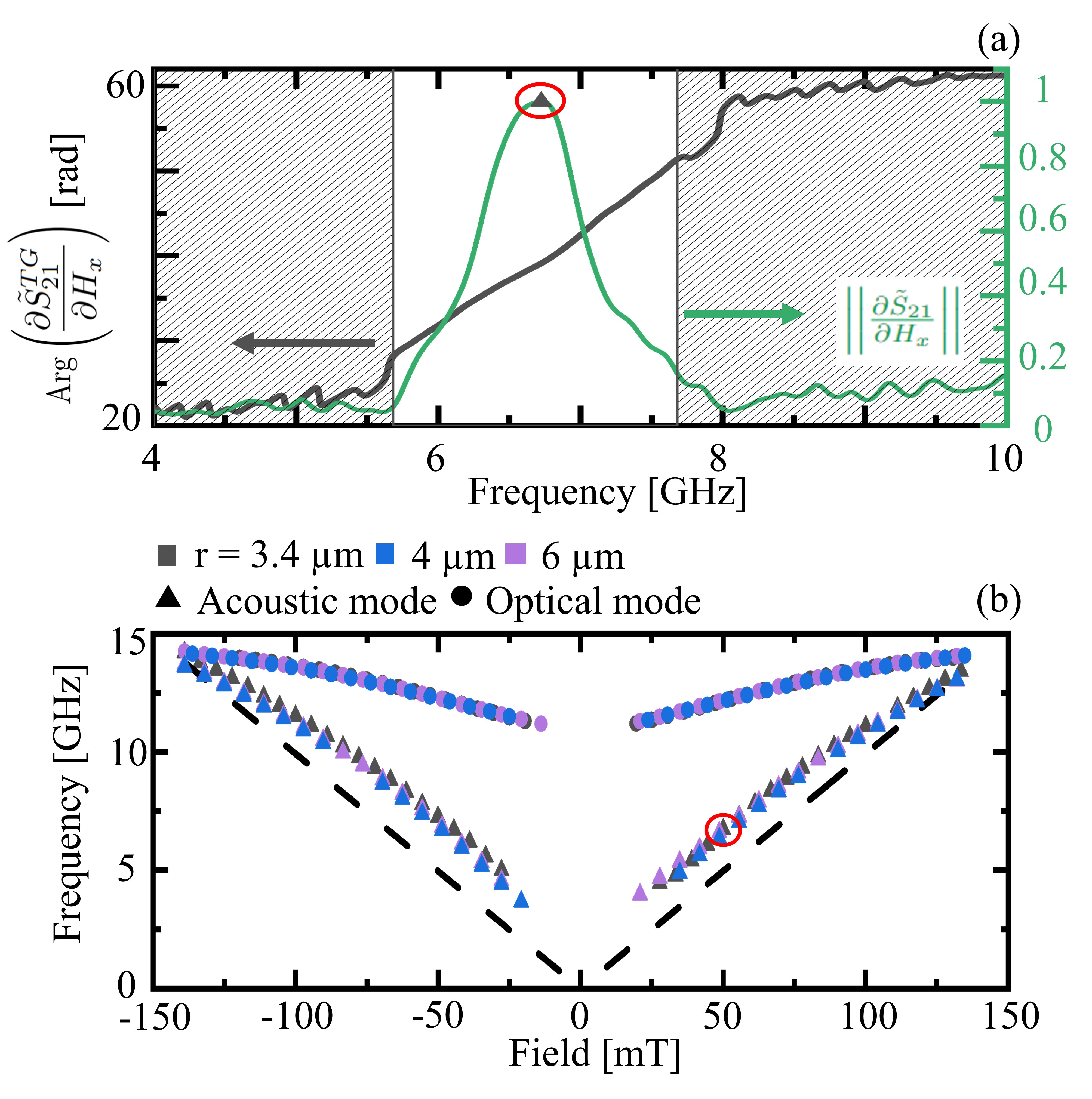}
    \caption{ (a) Phase (black) and modulus (green) of $\frac{\partial \tilde S_{21}}{\partial H_\textrm{x}}(f)$ for $\{r=3.4 ~\mu\textrm{m},~w_\textrm{ant}=1~\mu\textrm{m}\}$ at 50 mT. The SNR in the shaded areas prevent a reliable extraction of the phase. (b) Field-dependence of the frequency leading to a maximum of $ \Big\lvert\Big\lvert \frac{\partial \tilde S _{21}}{\partial H_x} \Big\rvert \Big\vert$ as function of the field for different propagation distances ($r$). For the acoustic (optical) mode, this corresponds exactly (approximately) to the uniform resonance frequency. The dashed lines are the prediction of the two-macrospin model \cite{devolder_measuring_2022}.}
    \label{fig:f(k=0)}
\end{figure}
Equations \ref{modulus} and~\ref{phase} show that it is essential to analyse both the modulus and the phase of field-derivative spectra in order to correctly determine the dispersion relation. Fig.~\ref{fig:f(k=0)}(a) displays a representative spectrum taken at 50 mT: The frequency of the acoustic mode at $k=0$ is easily found to be $f_0=6.7$ GHz where the signal-to-noise ratio is $\textrm{SNR} \approx10$. In the frequency region where the modulus is above the noise, the phase is reliably found to evolve rather monotonically. When the $\textrm{SNR} \leq 1$, the measurement noise prevents the estimate of the phase whose variations exhibit large fluctuations [shaded areas in Fig.~\ref{fig:f(k=0)}(a)]. The phase is reliable in a typically 2 GHz-wide frequency window.


The field dependence of the frequency at $k=0$, obtained with Eq.~\ref{modulus}, is displayed in Fig.~\ref{fig:f(k=0)} (b). The different propagation distances yield consistent estimates of ${\omega_0}/{(2 \pi)}$ with a maximum disagreement of $220$ MHz. The frequency of the uniform acoustic SW is essentially proportional to the field, in line with the behavior predicted by a 2-macropin model \cite{devolder_measuring_2022} (i.e., $\omega_0= \gamma \mu_0 H_x \sqrt{(M_s+H_j)/H_j} $, dashed line in Fig.~\ref{fig:f(k=0)}), with an additional curvature linked to the finite exchange stiffness within the CoFeB layers \cite{mouhoub_exchange_2023} and a finite gap related to their magnetic anisotropy \cite{seeger_inducing_2023}.


Quantitative estimates of $k(\omega)$ requires an estimate of the attenuation length. We use the estimated group velocity from the impulse response of Fig.~\ref{fig:signal_processing}(b) and the definition $L_\textrm{att}={2v_g}/{\Delta\omega_0}$. This yields a starting value of $L_\textrm{att}= 11.5~\pm 1.5~\mu$m, which is subsequently refined with further analysis. This $L_\textrm{att}$ results in effective propagation distances $r_\textrm{eff}$ that are slightly longer than the antenna-center-to-distances $r$ by $\{0.06, 0.2, 0.17\}~\mu$m for $r=\{3.4, 4, 6\}~\mu$m.

The dispersion relation is found by subtracting the phase at $f_0=  {\omega_0} /{(2 \pi)}$ and then dividing by $r_\textrm{eff}$ in the region of reliable phase extraction [e.g. in the non-shaded areas in Fig.~\ref{fig:f(k=0)}(a)],
 \begin{equation}
      k~=~\frac{1}{r_\textrm{eff}}\Big[\textrm{Arg} \left( \frac{\partial \tilde{S}_{21}(f)}{\partial H_x} \right)-\textrm{Arg} \left( \frac{\partial \tilde{S}_{21}(f_0)}{\partial H_x} \right)\Big]. \label{GettingOmegaOfk}
 \end{equation}

\begin{figure} 
\centering
\includegraphics[width=0.5\textwidth]{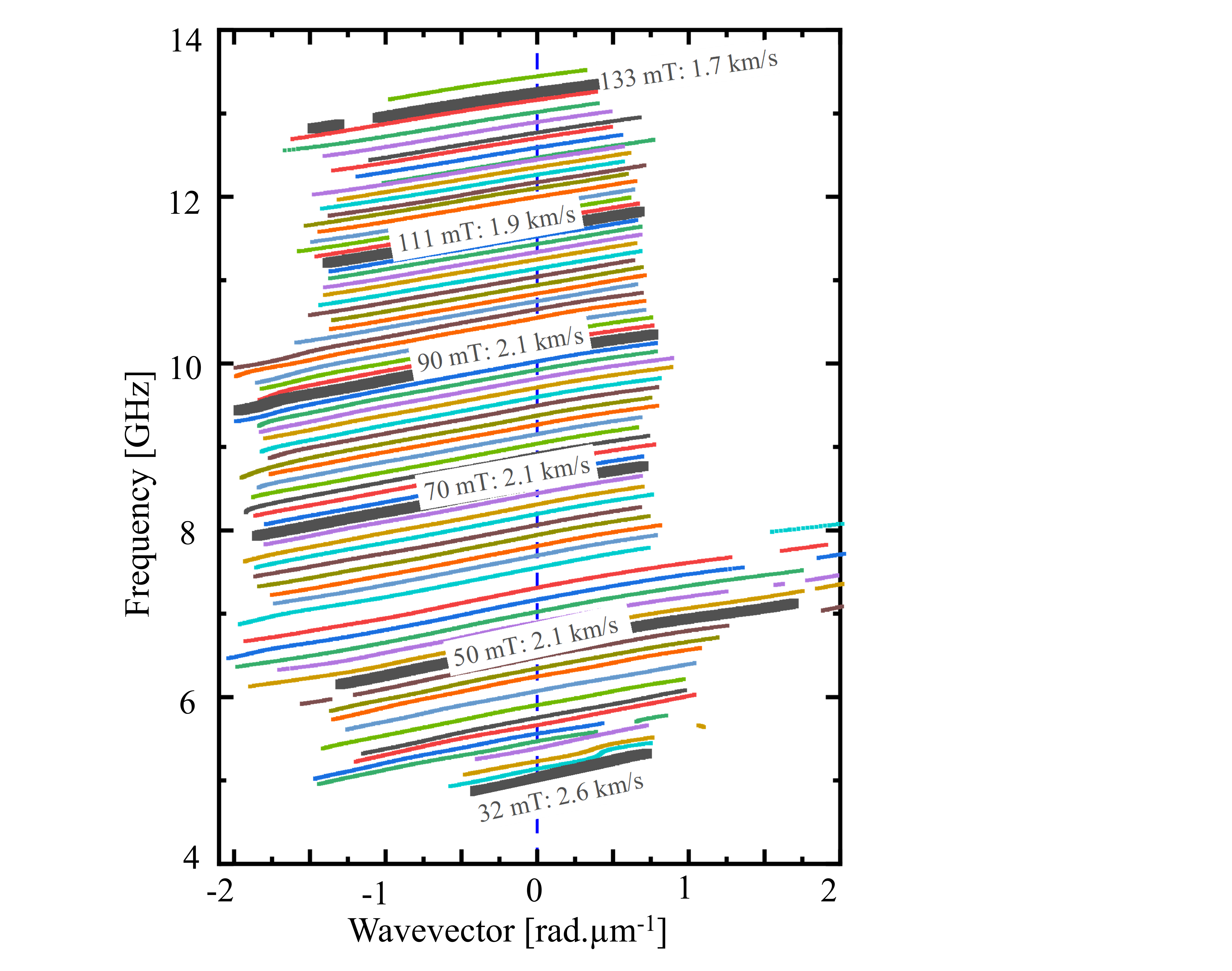}
\caption{ Experimental dispersion relations for applied fields ranging from 32 to 136 mT obtained from Eq.~\ref{GettingOmegaOfk} for a {device with $\{r, w_\textrm{ant}\}=\{4~\mu\textrm{m}, 2~\mu\textrm{m}\}$. The field is incremented by steps of $\approx$1.4 mT. The given velocities are obtained from the slopes of linear fits across the whole range of displayed wavevectors}}
\label{fig:DispRelAllFields}
\end{figure}

This procedure was implemented several times. In some field-frequency regions, measurement noise makes the obtained dispersion relations slightly different ; we have kept only the data points for which there is consensus within the ensemble of data \footnote{The 57 mT dataset has been removed because too noisy.}, such that the reliable $k$-points within the dispersion relations end up in being field-dependent (Fig.~\ref{fig:DispRelAllFields}), and sometimes sparse in wavevector space like between 47.8 and 54.2 mT. The unidirectional character of the spin waves is anyway clearly apparent: the group velocity is positive irrespective of the sign of the wavevectors.


\begin{figure} 
\centering
\includegraphics[width=0.45\textwidth]{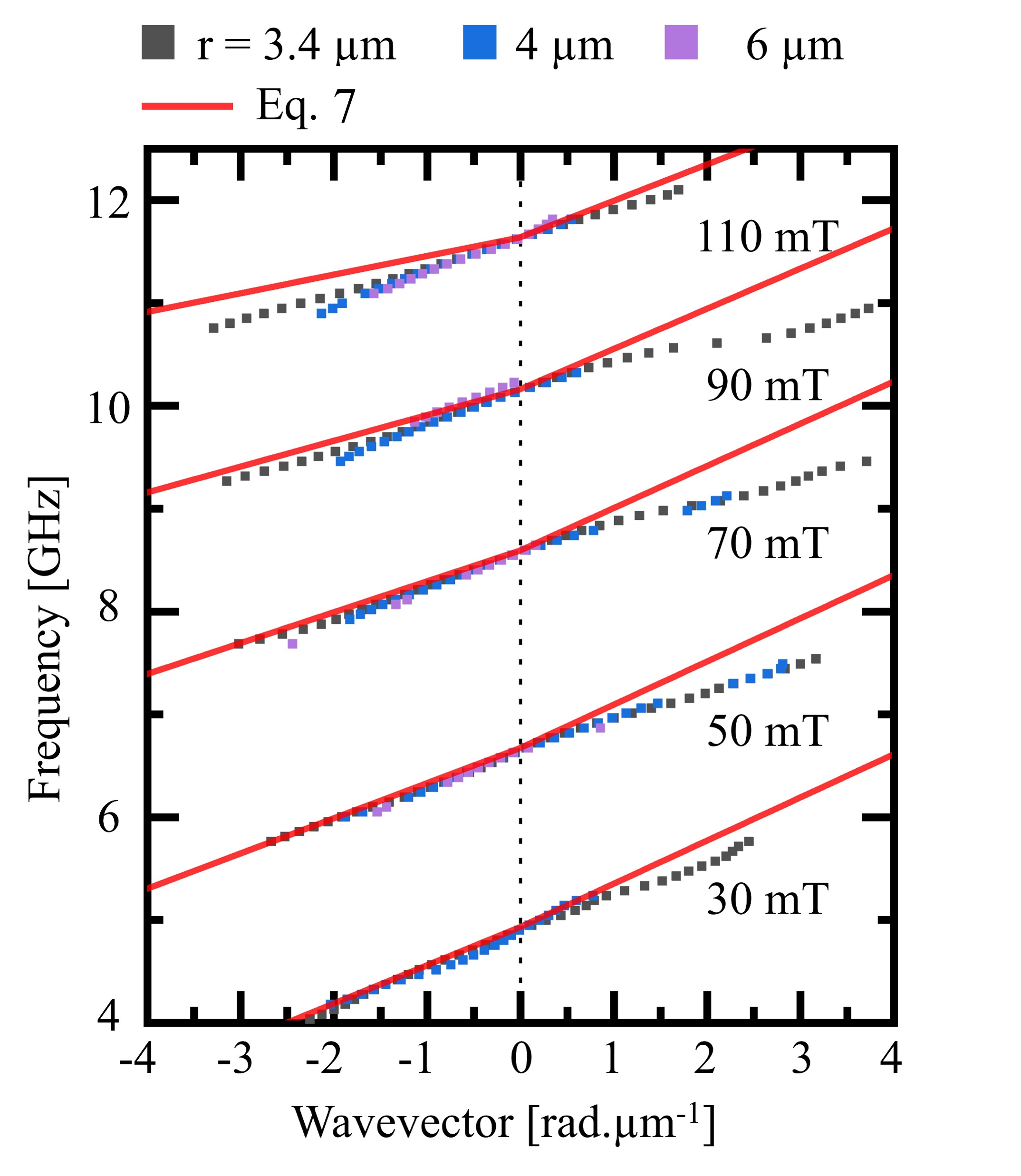}
\caption{ Comparison of the dispersion relations deduced from the measurements of devices with different antenna widths and separations. The red line is the prediction from Eq.~\ref{AnalyticalVg}. }
\label{fig:dispersion relation}
\end{figure}

The procedure to extract the dispersion relations from experimental data should obviously not depend on the propagation distance, nor on the antennas used. Fig.~\ref{fig:dispersion relation}.a, displays a comparison of $\omega(k)$ deduced from three different devices at selected applied fields, as well as $\omega(k) = \omega_0 + v_g k$ computed from the prediction  of ref.~\onlinecite{millo_unidirectionality_2023} which uses a sign-dependent group velocity,
\begin{equation}
v_g = \frac{\gamma_0 M_s t}{2  }  \left(\sqrt{1 - \left(\frac {H_x} {H_j}\right)^2} + \frac{|H_x| \textrm{sgn}(k)} {\sqrt{H_j (M_s+H_j)}} \right). \label{AnalyticalVg}
\end{equation}
The experimental dispersion relations are found to be almost identical with slopes differing by at most $\pm$6 \%.
The experimental dispersion relations are probed in a range of wavevectors that is naturally larger for narrower antennas (here: $w_\textrm{ant}=1~\mu\textrm{m}$). This range is also affected by the other geometrical parameters (more details are given in the Appendix \ref{appendix: antenna}). 
To a lesser extent, the k-range is also affected by the signal strength: shorter propagation distances yield  transmission signals that are usable up to slightly larger wavevectors.

As pointed in ref.~\onlinecite{millo_unidirectionality_2023}, Eq.~\ref{AnalyticalVg} accounts quantitatively for the dispersion relation only when the SAF ground state is well-described by two layers that are uniformly magnetized across their thickness. For thick SAFs like the films we studied, this translates in the condition $H_x \ll H_j$ \cite{mouhoub_exchange_2023}, which is satisfied only for the lowest applied fields investigated. 
For instance at 32 mT, Eq.~\ref{AnalyticalVg} would predict 2.58 km/s, while a linear fit through the experimental $\omega(k)$ yields 2.61 km/s. 

The disagreement between  Eq.~\ref{AnalyticalVg} and the experimental results becomes notably more pronounced as the applied field is increased. For the fields $\{50, 70, 90, 111, 133\}$ mT highlighted in Fig.~\ref{fig:dispersion relation}, the average slope of the experimental dispersion relations are $\{2.1, 2.1, 2.1, 1.9, 1.7\}$ km/s, while Eq.~\ref{AnalyticalVg} would predict $\{2.5, 2.3, 2.1, 1.7, 1.2\}$ km/s, i.e., a more pronounced reduction in the SW velocity with the applied field.

\section{Conclusion}
We have employed propagating spin wave spectroscopy to study the acoustic spin waves of a SAF when the spin wave wavevector is parallel to the applied field. The SW dispersion relation is found to be non-reciprocal and monotonic near $k=0$. We show that this monotonic character holds up to large applied fields. The standard semi-empirical procedure of analysing propagating spin wave spectroscopy experiments cannot be applied to analyse this situation. This stems from two difficulties: the $k=0$ spin wave frequencies are difficult to extract from the spectra, and the effective lengths of propagation of the spin waves are ill-defined. 

We developed an exact analysis technique by expanding the work reported in ref.~\onlinecite{devolder_propagating-spin-wave_2023}. The method relies on field-differentiation, time-of-flight filtering, and detailed analysis of the phase of the PSWS signal. We show in particular (Eq.~\ref{reff}) that since there are several characteristic lengths in the PSWS problem (antenna width $w_\textrm{ant}$, spacing $r$ between the antennas and spin attenuation length) the phase accumulated by the spin wave propagating between the two antennas is not strictly proportional to $r$,which contrasts with what the much-more studied case of even dispersion relations. Not accounting for the difference between the effective propagation distance and the antenna-to-antenna spacing can yield errors in estimating the wavevectors within the dispersion relations. The error is minimized by using as narrow as possible antennas, which also broadens the range of accessible wavevectors. The error is also minimized when using materials with long SW decay lengths, and devices with long propagation distances.

\begin{acknowledgments}
This work was supported by the French RENATECH network, by the French National Research Agency (ANR) as part of the “Investissements d’Avenir” and France 2030 programs. This includes the SPICY project of the Labex NanoSaclay: ANR-10-LABX-0035, the MAXSAW project ANR-20-CE24-0025 and the PEPR SPIN projects ANR 22 EXSP 0008 and ANR 22 EXSP 0004. We acknowledge Aurélie Solignac for material growth and Joo-Von Kim and Maryam Massouras for helpful comments.
\end{acknowledgments}

\appendix
\section{Forward versus reverse transmission parameters}
\begin{figure}[h!]
   \centering
    \includegraphics[width=0.5\textwidth]{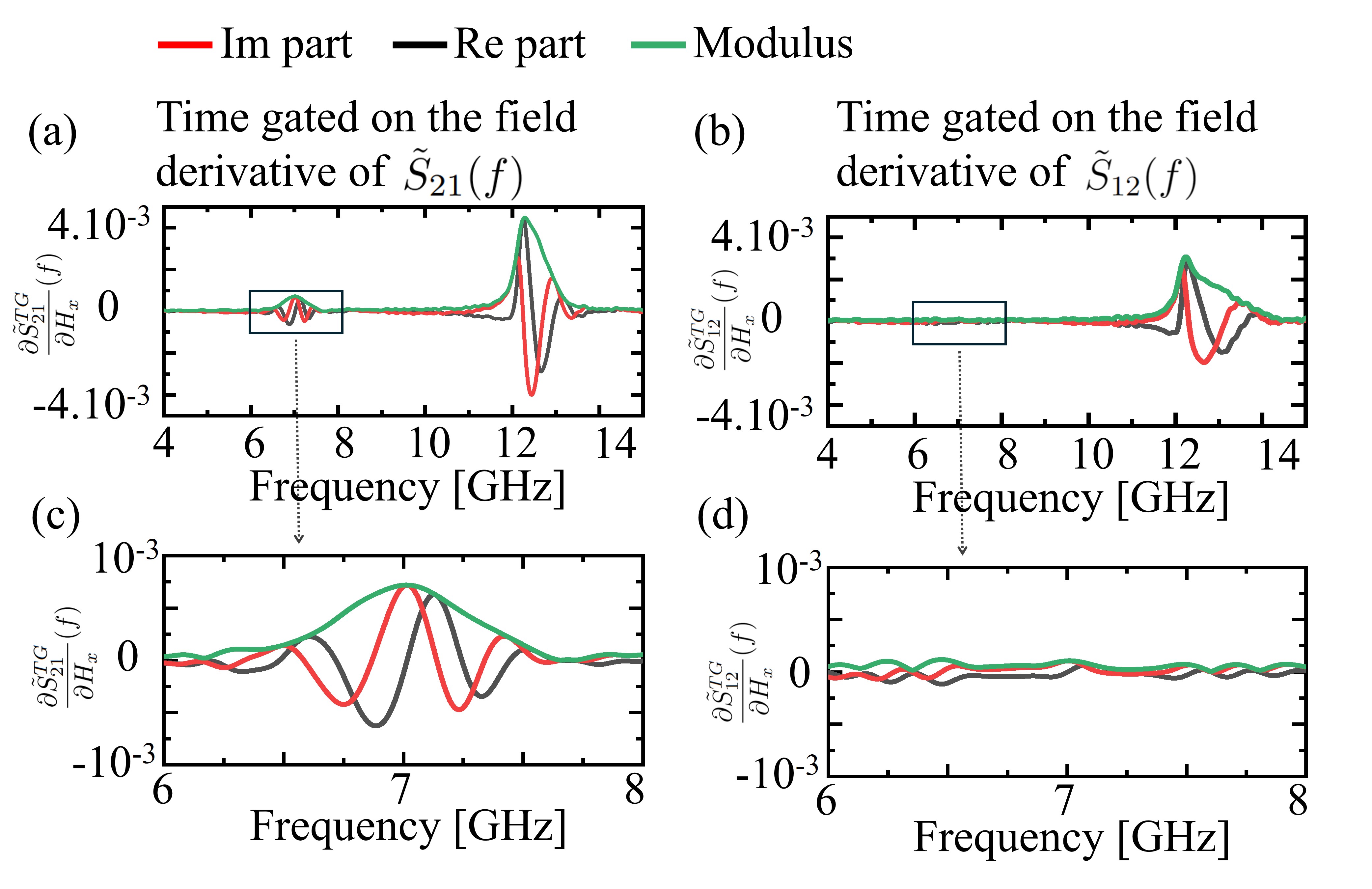}
    \caption{Spin wave spectra after various processing steps for $\mu_0H_x=-50~\textrm{mT}$, $w_\textrm{ant}=2~\mu$m and $r=4~\mu \textrm{m}$. (a) Time gated, field derivative of $\tilde S_{21}$.(b) Time gated, field derivative of $\tilde S_{12}$.(c-d) Zooms in the frequency region of the acoustic spin waves.}
    \label{fig:S21_vs_S12}
\end{figure}

The figures~\ref{fig:S21_vs_S12}(a) and (b) compare the forward and the backward transmission spectra. The contribution of the optical mode at $\approx12$ GHz is evident for both transmission directions. 
The contribution of the acoustic spin waves [Fig \ref{fig:S21_vs_S12} (c,d)] is perceivable in the $S_{21}$ spectrum, as clear oscillations at frequencies near 7 GHz. Conversely, no signal coming the acoustic spin waves can be perceived in the $\tilde S_{12}$ spectrum. It essentially consists of a noisy baseline near the acoustic frequency. 

This comparison shows the unidirectionality of the SAF acoustic spin waves, with finite transmission only in the forward direction when $\vec k \parallel \vec{H}_\textrm{applied}$.

\section{Dependence of the wavevector spectrum with the antenna geometry}
\label{appendix: antenna}
The range of wavevectors involved in propagating spin wave spectroscopy experiment depends largely on the antenna geometry: the antenna width $w_\textrm{ant}$, the antenna to film spacing $s$ and the antenna thickness $h$.
The field generated by a current $I$ passing through the antenna of rectangular cross section can be expressed as (Eq. 6, ref.~~\cite{devolder_propagating-spin-wave_2023}):
\begin{equation}
\label{eq:hrf}
    h_u^{rf}(k)=\frac{I}{2 \sqrt{2 \pi}} \textrm{sinc}(\frac{kw_\textrm{ant}}{2})\left(\frac{1-e^{-|k|h}}{h|k|} \right)e^{-|ks|}
\end{equation}
$h_u^{rf}(k)^2$ is commonly referred as the antenna efficiency function.

\begin{figure}[h!]
	\centering
	\includegraphics[width=0.5\textwidth]{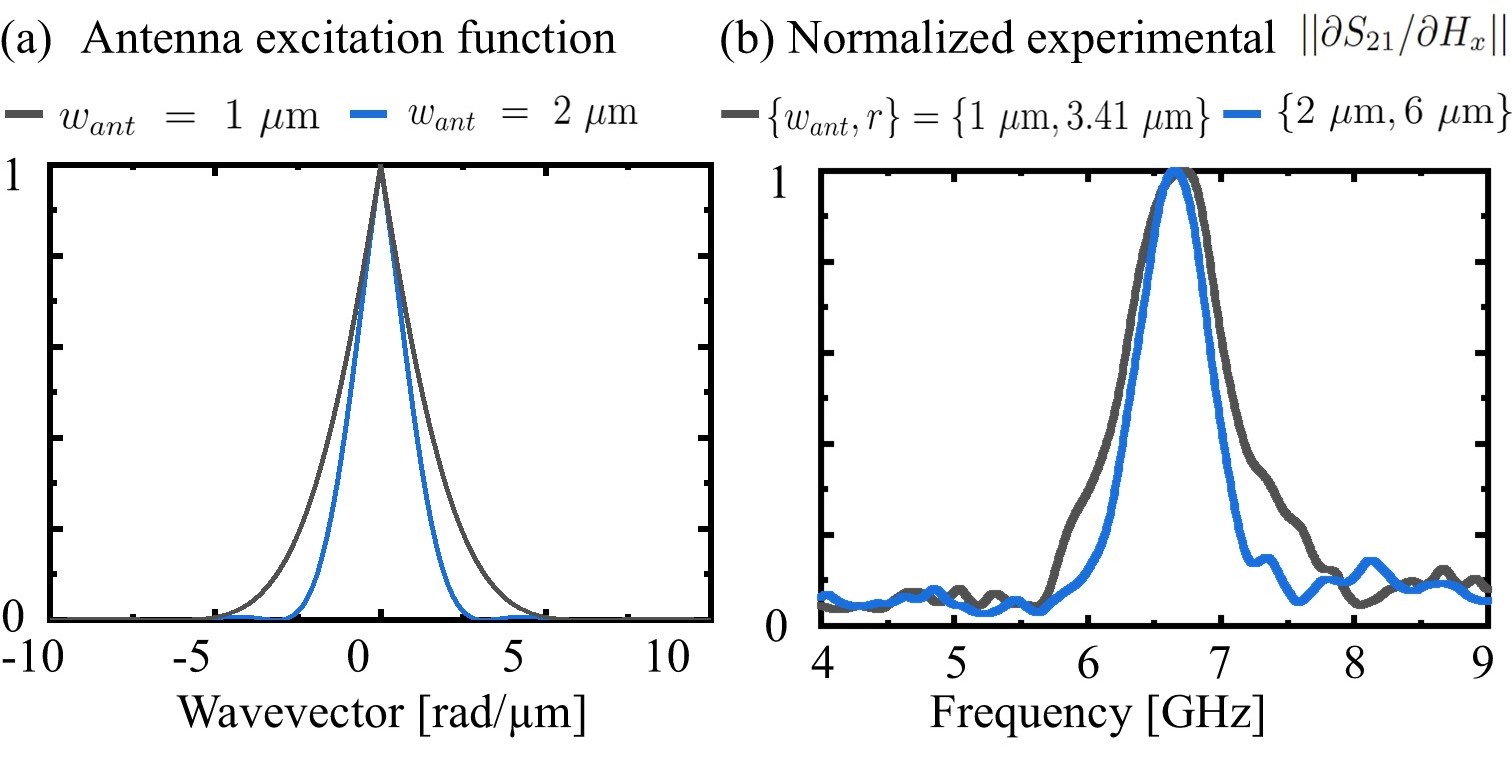}
	\caption{Antenna efficiency function compared to the modulus of the field derivative of the forward transmission parameter. (a) Wavevector dependence of the antenna efficiency for antenna width of $w_\textrm{ant}=~1~ \mu \textrm{m}$ and $w_\textrm{ant}=~2~ \mu \textrm{m}$. The antenna thickness and its distance from the magnetic film are 0.15 $\mu$m. (b) Experimental $ || \partial S_{21}/ \partial H_x ||$ after time-gating and for devices with the two considered antenna widths. The data are normalized to their maxima. 
   }
    \label{fig:antenna}
\end{figure}

Figure \ref{fig:antenna}(a) displays the antenna efficiency function in wavevector space for our antenna geometry (i.e. $s=h=0.15~\mu$m, and variable $w_\textrm{ant}$). The half width of the distribution is primarily set by $\pi/w_\textrm{ant}$, with a further reduction at finite $s$ and finite $h$.  
\\

For a quasi-linear dispersion relation $\omega \approx \omega_0 + v_g k $, this excitation efficiency in wavevector space is transferred to the frequency space within $|| \partial \tilde S_\textrm{21} / \partial H_x ||$ [Fig.~\ref{fig:antenna}(b)]. The center frequency of the signal is $\omega_0 / (2 \pi)$ and the frequency half width is now essentially 
$ 2 v_g /w_\textrm{ant}$. As a result, the frequency range where the signal is large is broader for the devices with a smaller antenna, just like the antenna efficiency function was, and the shapes of the two functions are very similar.

\newpage
\bibliography{SAFPSWS}

\end{document}